\documentclass[aps,pra,reprint,superscriptaddress,amsmath,amsfonts,showpacs]{revtex4-1}
\usepackage{bbm}
\usepackage{graphicx}

\newcommand{\ket}[1]{\left| #1 \right\rangle} 
\newcommand{\abs}[1]{\left\lvert {#1} \right\rvert} 
\newcommand{\op}[1]{\hat{#1}^{\mathop{\vphantom{\dagger}}}} 
\newcommand{\opd}[1]{\hat{#1}^\dagger} 
\newcommand{\mbeq}{\overset{!}{=}} 

\begin{document}

\title{Hong-Ou-Mandel interference without beam splitters}
 
\author{S.~M\"ahrlein}
\affiliation{Institut f\"{u}r Optik, Information und Photonik, Universit\"{a}t Erlangen-N\"{u}rnberg, 91058 Erlangen, Germany}
\affiliation{Erlangen Graduate School in Advanced Optical Technologies (SAOT), Universit\"at Erlangen-N\"urnberg, 91052 Erlangen, Germany}

\author{S.~Oppel}
\affiliation{Institut f\"{u}r Optik, Information und Photonik, Universit\"{a}t Erlangen-N\"{u}rnberg, 91058 Erlangen, Germany}
\affiliation{Erlangen Graduate School in Advanced Optical Technologies (SAOT), Universit\"at Erlangen-N\"urnberg, 91052 Erlangen, Germany}

\author{R.~Wiegner}
\affiliation{Institut f\"{u}r Optik, Information und Photonik, Universit\"{a}t Erlangen-N\"{u}rnberg, 91058 Erlangen, Germany}

\author{J.~von~Zanthier}
\affiliation{Institut f\"{u}r Optik, Information und Photonik, Universit\"{a}t Erlangen-N\"{u}rnberg, 91058 Erlangen, Germany}
\affiliation{Erlangen Graduate School in Advanced Optical Technologies (SAOT), Universit\"at Erlangen-N\"urnberg, 91052 Erlangen, Germany}

\date{\today}

\begin{abstract}
	We propose a new interferometric setup which displays a completely destructive generalized $N$-photon Hong-Ou-Mandel interference. The key property of this scheme is that is does not require any optical elements like beam splitters or integrated waveguide structures. The interference is intrinsically produced by the evolution of $N$ photons in free space when emitted by $N$ identical statistically independent single photon sources and measured by $N$ detectors in the far field. In this sense the setup is a most simple and natural implementation of the Hong-Ou-Mandel interference effect, i.e., of a completely destructive multi-photon interference produced by statistically independent incoherent sources.
\end{abstract}


\maketitle

\section{Introduction}
 
As first demonstrated by Hong, Ou and Mandel in 1987, two incoherent and statistically independent photons can interfere with each other if their modes are mixed by a beam splitter \cite{Hong1987,Shih1988}. More specifically, when sending the two photons  on a 50:50 beam splitter the number of coincident detection events at both output ports vanishes if the photons are indistinguishable with respect to polarization and frequency and the single photon wave packages overlap in time.
Such an interference effect manifests itself in the study of photon correlations rather than in the measurement of the intensity, and results from the completely destructive interference of the corresponding two-photon quantum paths.

The two-photon Hong-Ou-Mandel effect has been implemented with various kinds of single photon sources, e.g., single photons produced via spontaneous parametric down conversion (SPDC) in nonlinear crystals \cite{Hong1987,Shih1988}, or using ions \cite{Maunz2007}, atoms \cite{Grangier2006,Hofmann2012} and quantum dots as photon sources \cite{Patel2010, Hanson2012,Gold2014}. In these experiments the photons were mixed using a 50:50 beam splitter with two input and two output modes (see Fig.~\ref{fig:hom}). Besides systems with discrete optical elements the effect has also been studied in more complex setups such as integrated photonic circuits  including evanescently coupled waveguides \cite{Weihs1996,Politi2008,Rai2008,Bromberg2009} or coupled plasmonic systems \cite{Tame2013,Gupta2014,Martino2014,Fakonas2014}. The scheme has also been generalized to more involved network architectures with larger number of input photons \cite{Mattle1995,Campos2000,Lim2005,Buchleitner2010,Sanders2013,Tamma2014}, e.g., feeding three photons in a beam splitter arrangement or an integrated photonic device \cite{Wang2005a,Spagnolo2012,Spagnolo2013} and four photons in a fiber network or mixed by an array of unbalanced beam splitters \cite{Ou1999,Liu2007}. In recent years there has been particularly an increased interest in using integrated photonic waveguide structures where the waveguides are written with femtosecond lasers into solid materials or employing other methods \cite{Politi2008,Bromberg2009,Peruzzo2011a,Meany2012,Metcalf2013a,Spring2013,Broome2013,Crespi2013,Tillmann2013,Withford2014}, culminating in an integrated glass chip waveguide interferometer with 13 modes and three input photons \cite{Spagnolo2014}. However, in the mentioned generalized interferometers the number of input photons is usually much less than the number of in- and output modes. Moreover, for the integrated photonic circuits the manufacturing process still remains a challenge, even though there has been appreciable progress recently \cite{Tanzilli2012}. In particular, the waveguide arrays are static so that the circuit parameters can not be changed easily, i.e., for a new set of parameters a new integrated waveguide structure has to be manufactured. 

In contrast to elaborate waveguide structures we discuss in this paper a very simple and versatile Hong-Ou-Mandel interferometer where an arbitrary number of photons are fed into an equal number of input modes and coincidentally measured in the same number of output modes. In the interferometer no optical elements like beam splitters or integrated waveguide structures are used. More specifically, the interferometer relies on $N$ photons emitted by $N$ identical statistically independent incoherent single photon sources and measured by $N$ detectors located in the far field. The interferometer requires only identical single photon sources and single photon detectors recording the photons in free space. As a consequence, experimental difficulties such as phase instabilities or mode mismatch are strongly reduced. Since the photons propagate in free space we can also ensure minimal losses due to absorption along the paths towards the detectors. As it turns out, due to its design, $N$-photon interference is an intrinsic property of the setup, i.e., the $N$-photon interference has not to be engineered by any passive or active optical elements. In this sense we propose a most simple realization of the $N$-photon Hong-Ou-Mandel interferometer, displaying a completely destructive interference for particular detector configurations.

The paper is organized as follows. We start with a short revision of the original Hong-Ou-Mandel two-photon interference experiment in Sec.~\ref{sec:HOM}. In Sec.~\ref{sec:2freespace} we present our free space Hong-Ou-Mandel interferometer for two input photons, displaying a completely destructive two-photon interference in correspondence with the original Hong-Ou-Mandel experiment. In Sec.~\ref{sec:Nfreespace} we generalize our scheme to $N$ input photons and demonstrate that this $N$-photon interferometer displays a completely destructive $N$-photon Hong-Ou-Mandel interference as well. In Sec.~\ref{sec:Nfreespace} we present our conclusions.

\section{Original Hong-Ou-Mandel experiment \label{sec:HOM}}

\begin{figure}
	\centering
	\includegraphics{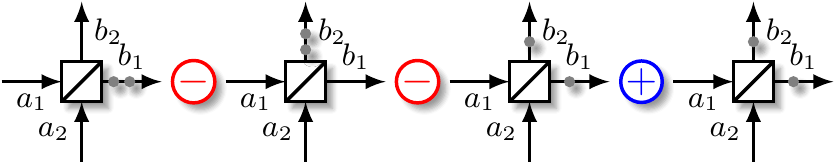}
	\caption{Two-photon Hong-Ou-Mandel effect. The mixing of the two input modes $a_{1}$ and $a_{2}$ by a $50:50$ beam splitter leads to a complete destructive interference of the output configuration $\opd{b}_1 \opd{b}_2 \ket{0}$. This means that two photons impinging on the beam splitter in mode $a_{1}$ and $a_{2}$ will always leave the beam splitter in the same mode, either both in mode $b_1$ or both in mode $b_2$, but never separately in the two modes $b_{1}$ and $b_{2}$ at the same time. \label{fig:hom}}
\end{figure}

In the original Hong-Ou-Mandel experiment \cite{Hong1987} two incoherent photons are created via SPDC, sent in two different input modes onto a 50:50 beam splitter, and the number of two-photon coincidences is measured at the two output ports. If the photons are indistinguishable with respect to frequency, polarization and time, the number of coincidences drops to zero. The two-photon interference effect manifests itself in the coincident second order spatial intensity correlation function $G^{(2)}(\vec{r}_1, \vec{r}_2)$ at the output ports, where \cite{Glauber(1963)Quantum}
\begin{equation}
\begin{split}
\label{eq:secondordercor}
G^{(2)}(\vec{r}_1, \vec{r}_2) &\sim \left\langle \opd{b} (\vec{r}_1) \opd{b} (\vec{r}_2)  \op{b} (\vec{r}_2)  \op{b} (\vec{r}_1) \right\rangle_{\ket{\psi_\text{out}}} \\
& \equiv \left\langle \opd{b}_1 \opd{b}_2  \op{b}_2 \op{b}_1 \right\rangle_{\ket{\psi_\text{out}}} \; ,
\end{split}
\end{equation}
and $\op{b} (\vec{r}_m) = \op{b}_m$ ($\opd{b} (\vec{r}_m) = \opd{b}_m$), $m = 1, 2$, denotes the annihilation (creation) operator of a photon in the output mode $b_m$. 

To determine the evolution of the creation and annihilation operators for the two incoming photons we note that $\opd{a}_{1}$ and $\opd{a}_{2}$ are transformed by a symmetric 50:50 beam splitter into the creation operators of the output modes $\opd{b}_{1}$ and $\opd{b}_{2}$ via \cite{Agarwal2012}
\begin{equation}
	\begin{pmatrix}\opd{a}_1 \\ \opd{a}_2\end{pmatrix} \rightarrow \frac{1}{\sqrt{2}} \begin{pmatrix} 1 & 1\\ 1 & -1 \end{pmatrix} \begin{pmatrix}\opd{b}_1 \\ \opd{b}_2\end{pmatrix} \, .
	\label{eq:BS}
\end{equation}
This means that if we start with an initial state $\ket{\psi_\text{in}}$ that has a single photon in each input mode, the beam splitter transforms this state at the output in the following way
\begin{equation}
\begin{split}
\label{eq:evoluop1}
	\ket{\psi_\text{in}} = & \;  \opd{a}_1 \opd{a}_2 \ket{0} = \ket{1,1} \\ &\rightarrow \ket{\psi_\text{out}} = \frac{1}{2} \left( (\opd{b}_1)^2  - (\opd{b}_2)^2 - \opd{b}_1 \opd{b}_2 + \opd{b}_2 \opd{b}_1  \right) \ket{0} \, .
\end{split}
\end{equation}
As photons are bosons, we have to consider the commutator relation $\left[ \opd{b}_j, \opd{b}_k \right] = 0$, what yields
\begin{equation}
\label{eq:evoluop2}
	\ket{\psi_\text{in}} = \ket{1,1} \rightarrow \ket{\psi_\text{out}}=\frac{1}{\sqrt{2}} \left( \ket{2,0} - \ket{0,2} \right) \, .
\end{equation}
One can see that in the output state no term $\propto \ket{1,1}$ is present and consequently $G^{(2)}(\vec{r}_1, \vec{r}_2)$ drops to zero. This means that both photons will always leave the beam splitter in the same mode, either both in mode $b_1$ or both in mode $b_2$, but never separately in the two modes $b_1$ and $b_2$ at the same time. Consequently, no coincident two-photon detection events are observable at the two output modes. We stress that in the considered setup the Hong-Ou-Mandel interference effect arises from the (unitary) bosonic mode mixing by the beam splitter, as displayed by Eqs.~(\ref{eq:BS}) - (\ref{eq:evoluop2}). A graphical interpretation of the experiment is sketched in Fig.~\ref{fig:hom}.

\section{Two-photon Hong-Ou-Mandel interference in free space \label{sec:2freespace}}
\label{sec:three}

\subsection{Experimental setup}

\begin{figure}
	\centering
	\includegraphics{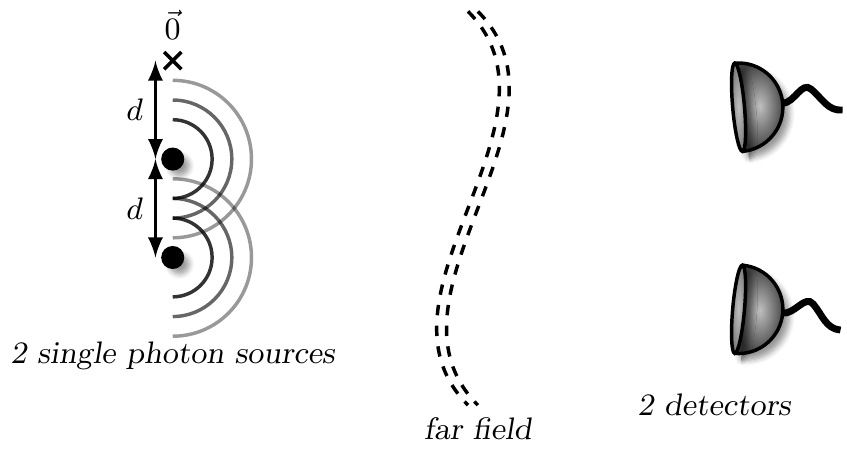}
	\caption{Scheme for an experimental realization of the two-photon Hong-Ou-Mandel interference effect in free space \label{fig:2photonsetup}.}
\end{figure}

Next we investigate an experimental setup which implements the operation Eq.~(\ref{eq:BS}) in free space. To this end we consider the scheme sketched in Fig.~\ref{fig:2photonsetup}. Here, two identical statistically independent single photon sources are located at positions $\vec{R}_1$ and $\vec{R}_2$ along the y-axis, separated by a distance $d$ assumed to be large enough that any coupling between the sources can be neglected. To simplify the calculations, we choose the origin of the coordinate system a distance $d$ off the first source (see Fig.~\ref{fig:2photonsetup}). The single photon emitters can be realized by, e.g., two identical 2-level atoms with upper state $\ket{e}_n$ and ground state $\ket{g}_n$, $n = 1, 2$, excited initially to the upper level. Via spontaneous decay the two atoms relax into the ground state, thereby emitting each a single photon of equal polarization and frequency. Since the photons have the same wavelength $\lambda$ the corresponding wave vectors have the same norm and differ only in their orientation.

In order to determine the spatial correlations between the photons in the far field of the atoms, we use two detectors placed at positions $\vec{r}_1$ and $\vec{r}_2$, with $| \vec{r}_1 | = | \vec{r}_2 | >> d$, and measure the second order spatial intensity correlation function $G^{(2)}(\vec{r}_1, \vec{r}_2)$. The far field condition is vital to this scheme since it leads to a loss of ``which-way'' information of the individual photons: the wave vectors of the two photons pointing in the far field towards the same detector are identical and hence are associated with the same ``plane wave mode''.  This means if one of the two photons is measured by a detector at $\vec{r}_m$, $m = 1,2$ , in the far field, there is no possibility of telling by which of the two sources this recorded photon was initially produced. 

\subsection{Emission by atoms: relation between ``spherical modes'' and ``plane wave modes''}

\begin{figure}
	\centering
	\includegraphics{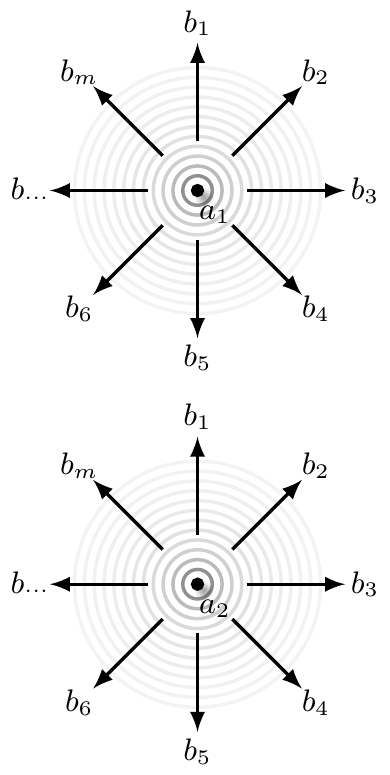}
	\caption{Relation between ``spherical modes'' $a_n$ and ``plane wave modes'' $b_m$.}
	\label{fig:modes}
\end{figure}

We employ a similar approach as in \cite{Wiegner2010} to illustrate how the excited atoms emit a single photon. As depicted in Fig.~\ref{fig:modes} we can describe a photon by either being in  a ``spherical mode'' $a_n$ or being in  a ``plane wave mode'' $b_m$.  Here, the expression ``spherical mode'' is used in the sense that $\opd{a}_n$ creates a photon in a spherical wave mode originating at $\vec{R}_n$, whereas $\opd{b}_m$ creates a photon in a plane wave mode with wave vector $\vec{k}_m$. Note that principally a two-level atom would radiate according to a three-dimensional dipole mode pattern. However, in the following we assume that the atoms and detectors are in one plane and that the atomic dipoles are oriented perpendicular to this plane. In this case the atoms emit isotropically within the detection plane and we can treat the spontaneous emission as occurring in all directions with equal probability (see Figs.~2 and 3). 

In general the evolution via spontaneous decay from a single two level atom in the excited state $\ket{e}$ and no photons in any mode (i.e. the vacuum state) to the atom in the ground state $\ket{g}$ and a single photon potentially in every possible mode $b_m$ can be described by
\begin{equation}
	\ket{e,0} \rightarrow \sum_{m} c_{m} \ket{g,1_{b_m}} \, ,
\label{eq:evolution}
\end{equation}
where the sum runs over all possible (i.e. an infinite number of) modes $b_m$and the coefficients $c_m$ can be obtained from Weisskopf-Wigner theory \cite{Scully}.

If more than one atom is involved, the phase factors of the coefficients $c_{nm} \propto e^{ - \text{i} \, \vec{R}_n \cdot \vec{k}_m}$ have to be taken into account. Here, the coefficients $c_{nm}$ refer to a photon produced by the $n$-th atom at $\vec{R}_n$ and populating a plane wave mode with wave vector $\vec{k}_m$. The system can thus be described by transforming the operators $\opd{a}_n$, which create one photon in a ``spherical mode'' at the coordinate $\vec{R}_n$, to the set of modes $\opd{b}_m$, which create a photon in the direction of $\vec{k}_m$ (see Fig.~3)
\begin{equation}
	\opd{a}_n \rightarrow \sum_{m} c_{nm} \, \opd{b}_m \, .
	\label{eq:transformation}
\end{equation}

In the following we imply the far field condition, i.e., photons emitted in the same direction by different atoms can not be distinguished and hence populate the same ``plane wave mode''. In this way we can achieve a mode mixing without optical elements. 

In the general case there would be an infinite number of modes $a_n$ (as there is an infinite number of positions) and an infinite number of modes $b_m$ (as there is an infinite number of directions). This would correspond to an infinite number of coefficients $c_{nm}$, i.e., an infinite transformation matrix $\left\{c_{nm}\right\}$. However, as we restrict our setup to two atoms at fixed positions ${\vec{R}_1}$ and ${\vec{R}_2}$, we limit the ``spherical modes'' to $a_1$ and $a_2$. Also since we only consider coincidence measurements at $\vec{r_1}$ and $\vec{r_2}$ the same is true for the ``plane wave modes'', i.e., we only need to take into account the modes $b_1$ and $b_2$, populated by the two photons before being recorded by the two detectors at ${\vec{r}}_1$ and ${\vec{r}}_2$, respectively.
This creates a $2\times2$ submatrix of $\left\{c_{nm}\right\}$, where we associate the rows with the positions of the two atoms and the columns with the wave vectors pointing towards the two detectors. As we assume that all directions in which photons could be emitted by the atoms are equally probable, the absolute value of all coefficients is given by the same normalization factor $c_\text{norm}$. Nevertheless, each coefficient is still characterized by certain relative phase factors depending on the position of the atom and the direction of the emitted photon, i.e.,
\begin{equation}
	c_{nm} = c_\text{norm} e^{ - \text{i} \, \vec{R}_n \cdot \vec{k}_m} \, ,
\label{eq:cnm}
\end{equation}
where $n=1,2$ labels the position of the photon sources and $m=1,2$ the position of the detectors.

\subsection{Geometry of the setup, path differences and phase factors}
\label{sec:path-phase}

\begin{figure}
	\centering
	\includegraphics{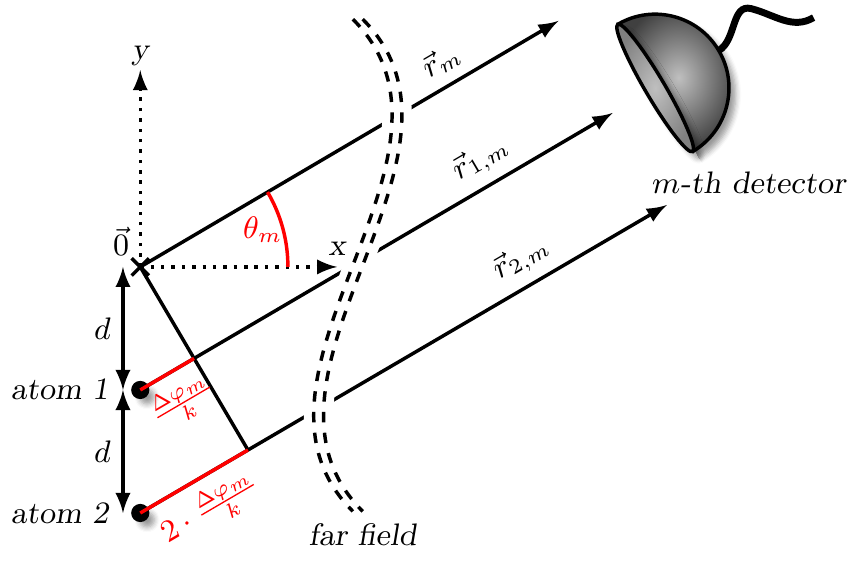}
	\caption{Origin of the phase difference between the single-photon quantum paths leading from atom $n = 1, 2$ to the m-th detector at $\vec{r}_m$, given by $\varphi_{n,m} = n \cdot \Delta \varphi_m$ (see Eq.~(\ref{eq:phasenm})).}
	\label{fig:2photongeometry}
\end{figure}

The phase factor $e^{ - \text{i} \, \vec{R}_n \cdot \vec{k}_m}$ of Eq.~(\ref{eq:cnm}) is due to the phase accumulated by a photon emitted from the $n$-th source and recorded by the $m$-th detector relative to a photon emitted at the origin and recorded by the same detector (see Fig.~\ref{fig:2photongeometry}). Since the sources are located along the y-axis and separated by a distance $d$, the relative phase of the photons emitted by the two atoms and recorded by the $m$-th detector is given by
\begin{equation}
	\Delta \varphi_m = k \cdot d \cdot \sin(\theta_m) \, ,
\label{eq:phasedifference}
\end{equation}
where $\theta_m$ is the angle between the wave vector $\vec{k}_m$ pointing towards the $m$-th detector and the x-axis (see Fig.~\ref{fig:2photongeometry}). 
From this it is easy to see that the relative phase of each path from source $n$ to detector $m$ given by
\begin{equation}
	\varphi_{n,m} = n \cdot \Delta \varphi_m \, .
\label{eq:phasenm}
\end{equation}
Note that by changing the position of the two detectors we can thus choose any value for $\Delta \varphi_{1}$ and $\Delta \varphi_{2}$ between 0 and $2 \pi$ (or to be exact between $-k \cdot d$ and $k \cdot d$).

\subsection{Two-photon Hong-Ou-Mandel interference}

\begin{figure}
	\centering
	\includegraphics{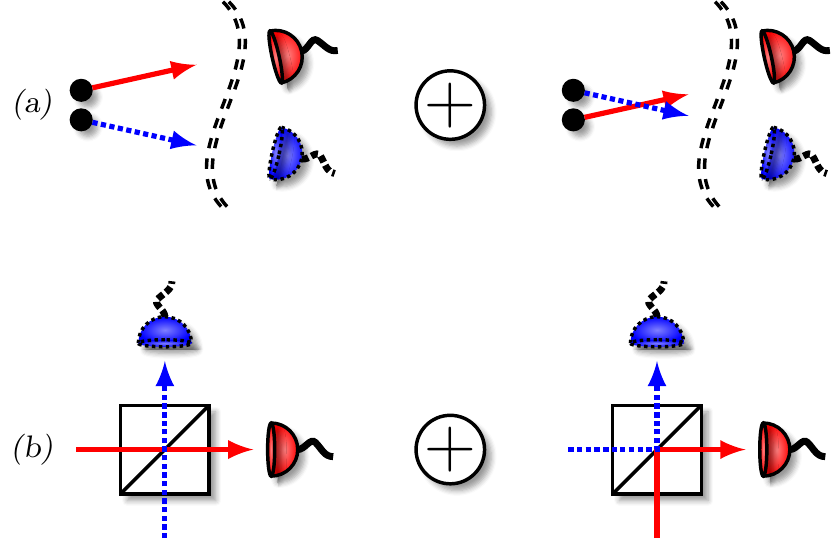}
	\caption{(Color online) The two indistinguishable two-photon quantum paths leading to a coincident measurement $G^{(2)}(\vec{r}_1, \vec{r}_2)$ of two photons (a) for the free space setup and (b) for the original Hong-Ou-Mandel setup employing a beam splitter. In both cases the two possible quantum paths are: (left) source 1 emits a photon recorded at detector 1 (red, solid) and source 2 emits a photon recorded at detector 2 (blue, dotted); (right) source 1 emits a photon recorded at detector 2 (blue, dotted) and source 2 emits a photon recorded detector 1 (red, solid).}
	\label{fig:path interference}
\end{figure}

The phase factors of the coefficients $c_{nm}$ of the transformation matrix produce interferences among different quantum paths leading to a given measurement configuration for the second order correlation function $G^{(2)}(\vec{r}_1, \vec{r}_2)$. To illustrate this we consider a coincidence measurement of two photons at $\vec{r}_1$ and $\vec{r}_2$, as depicted in Fig.~\ref{fig:path interference} (a). As the measurement is done in the far field we can not discriminate between the two two-photon quantum paths depicted in Fig.~\ref{fig:path interference} (a), namely source 1 emits a photon recorded at detector 1 and source 2 emits a photon recorded at detector 2, as well as, source 1 emits a photon recorded at detector 2 and source 2 emits a photon recorded at detector 1. Both two-photon quantum paths lead to the same final state, i.e., the same coincident measurement $G^{(2)}(\vec{r}_1, \vec{r}_2)$ of two photons at the two detectors. In order to calculate $G^{(2)}(\vec{r}_1, \vec{r}_2)$ we thus have to take into account the quantum mechanical superposition of these two two-photon quantum paths, i.e.,
\begin{equation}
\begin{split}
G^{(2)}(\vec{r}_1, \vec{r}_2) &\sim \abs{e^{-\text{i} \left( \varphi_{1,1} + \varphi_{2,2} \right)} +  e^{-\text{i} \left( \varphi_{1,2} + \varphi_{2,1} \right)}}^2 \\
 &= \abs{e^{-\text{i} \left( \Delta \varphi_1 + 2 \Delta \varphi_2 \right)} + e^{-\text{i} \left( \Delta \varphi_2 + 2 \Delta \varphi_1 \right)}}^2 \\
&= 2 \big( 1 + \cos(\Delta \varphi_1 - \Delta \varphi_2) \big) \, . 
\label{eq:2pathinterference}
\end{split}
\end{equation}
One can easily see that for $\Delta \varphi_1 = 0$ and $\Delta \varphi_2 = \pi$ the two two-photon quantum paths interfere completely destructively so that the probability of measuring a coincidence event is zero. This is the Hong-Ou-Mandel effect in free space, implemented without optical elements. It can be understood in direct analogy to the original Hong-Ou-Mandel experiment, where the two two-photon quantum paths leading to a coincident detection of the two photons cancel each other out (see Sec.~\ref{sec:HOM} and Fig. \ref{fig:path interference} (b)). Note that the Hong-Ou-Mandel interference in free space only occurs at certain detector positions -  a degree of freedom, which does not exist in the original experiment, where the beam splitter implements merely the fixed phase relation $\Delta \varphi_1 - \Delta \varphi_2 = \pi$. Note furthermore, that $\Delta \varphi_1 = 0$ and $\Delta \varphi_2 = \pi$ is not the only possible detector configuration to observe $G^{(2)}(\vec{r}_1, \vec{r}_2) = 0$. More generally, the condition $\Delta \varphi_1 - \Delta \varphi_2 = (2j+1) \pi$ with $j \in \mathbb{Z}$ has to be fulfilled.

\section{$N$-photon Hong-Ou-Mandel interference in free space \label{sec:Nfreespace}}
\label{sec:four}

Finally, we discuss the generalization of the two-photon Hong-Ou-Mandel interferometer in free space to an $N$-photon Hong-Ou-Mandel interferometer in free space with $N$ input photons and $N$ output photons. To this end we extend the scheme introduced in Sec.~\ref{sec:three} to an arbitrary number $N$ of sources and detectors.

\subsection{Basic setup and measurement procedure}

\begin{figure}
	\centering
	\includegraphics{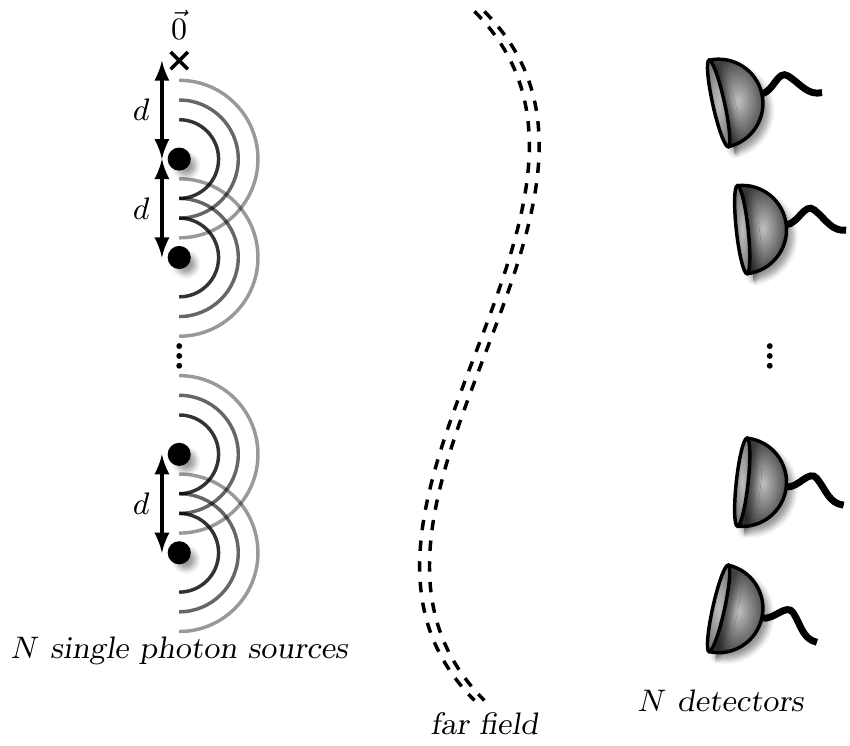}
	\caption{Scheme to realize the generalized $N$-photon Hong-Ou-Mandel effect in free space.}
	\label{fig:setup}
\end{figure}

A sketch of the setup is shown in Fig.~\ref{fig:setup}. We consider $N$ identical statistically independent single photon sources, equidistantly arranged along the y-axis at positions $\vec{R}_n$, $n = 1, \ldots, N$, with equal distance $d >> \lambda$ between the sources such that any interaction between the sources can be neglected. The origin of the coordinate system is assumed to be a distance $d$ off the first source along the y-axis. As in Sec.~\ref{sec:2freespace} we imagine the single photon sources to be realized by 2-level atoms, with ground state $\ket{g}_n$ and excited state $\ket{e}_n$, $n = 1, \ldots, N$, where all atoms are assumed to be initially excited to the upper level. Via spontaneous emission the atoms decay into the ground state, thereby emitting each a single photon. In order to measure the $N$-th order spatial intensity correlation function $G^{(N)}(\vec{r}_1, \ldots, \vec{r}_N)$ we employ $N$ detectors located in the far field of the sources. As before, the far field condition is vital to the scheme as it ensures the loss of ``which-way'' information of the individual photons. 

\subsection{Geometry of the setup, path differences and phase factors \label{sec:path-phase2}}
\label{subsec:Ngeo}

Analogously to Sec.~\ref{sec:three}, we employ the Weisskopf-Wigner theory \cite{Scully,Wiegner2010} to describe the emission of a photon by an atom via spontaneous decay. A photon produced in a ``spherical mode'' by the $n$-th atom at position $\vec{R}_n$ (implemented by the operator $\opd{a}_n$) is scattered into all possible ``plane wave modes'' with corresponding wave vectors $\vec{k}_m$ (described by the operator $\opd{b}_m$), i.e., we have the transformation (see Eq.~(\ref{eq:transformation}) and Fig.~\ref{fig:modes})
\begin{equation}
\label{eq:trafoone}
	\opd{a}_n \rightarrow \sum_{m} c_{nm} \, \opd{b}_{m} \, ,
\end{equation}
where $c_{nm} = c_\text{norm} e^{ - \text{i} \, \vec{R}_n \cdot \vec{k}_m}$ is the coefficient of the transformation matrix $\left\{ c_{nm} \right\}$, taking into account the (relative) optical path accumulated by a photon emitted at $\vec{R}_n$ and recorded at $\vec{r}_m$. As we assume that all directions in which the photons can be emitted are equally probable, the absolute value of all coefficients is given by the same real normalization factor $c_\text{norm}$. 

As the $m$-th detector is located in the far field, all vectors $\vec{r}_{n,m}$ pointing from the different atoms at $\vec{R}_n$, $n = 1, \ldots, N$, towards the same detector at $\vec{r}_m$, are aligned parallel and hence can be associated with the same ``plane wave mode''. This means, that if a photon is recorded by a detector at $\vec{r}_m$ there is no possibility to tell from which atom the photon was initially emitted (see Sec.~\ref{sec:three}). This implements the mode mixing of the photons in free space without optical elements. 

As the $N$ atoms are assumed to be equidistantly aligned along the y-axis, the phase accumulated by a photon emitted from the $n$-th atom at $\vec{R}_n$ and recorded by the $m$-th detector at $\vec{r}_m$ relative to a photon emitted at the origin is given by
\begin{equation}
	\varphi_{n,m} = n \cdot \Delta \varphi_m \, ,
	\label{eq:phasedifference2}
\end{equation}
where $\Delta \varphi_m$ corresponds to the phase difference between photons scattered by adjacent atoms 
\begin{equation}
\label{phasetodetector}
	\Delta \varphi_m = k \cdot d \cdot \sin(\theta_m) \, .
\end{equation}
Again $\Delta \varphi_m$ can assume any value between 0 and $2 \pi$ (or to be exact between $-k \cdot d$ and $k \cdot d$). In this way we can rewrite the coefficients $c_{nm}$ of the transformation matrix as 
\begin{equation}
	c_{nm} = c_\text{norm} e^{ - \text{i} \, n \cdot \Delta \varphi_m} \, .
	\label{eq:cnm2}
\end{equation}

Considering the emission of $N$ photons by $N$ atoms at positions $\vec{R}_1, \ldots, \vec{R}_N$, implemented by the operators $\opd{a}_1, \ldots ,\opd{a}_N$, the corresponding input state is given by
\begin{equation}
	\ket{\psi_\text{in}} = \prod_{n=1}^{N} \opd{a}_n \ket{0} \, .
	\label{eq:2input}
\end{equation}
According to Eq.~(\ref{eq:trafoone}) the transformation from the ``spherical modes'' $a_n$ to the ``plane wave modes'' $b_m$ reads
\begin{equation}
	\ket{\psi_\text{out}} = \prod_{n=1}^{N} \sum_{m} c_{nm} \, \opd{b}_m \ket{0} \, ,
	\label{eq:2output}
\end{equation}
where the sum runs over all possible directions $m$. Note that again this transformation involves principally an infinite number of ``plane wave modes'' $b_m$ and hence leads to an infinite transformation matrix $\left\{ c_{nm} \right\}$. However, since we only consider measurements $G^{(N)}(\vec{r}_1, \ldots, \vec{r}_N)$, where the $N$ photons are recorded coincidentally by $N$ detectors at $\vec{r}_m$, $m = 1, \ldots, N$, the infinite transformation matrix $\left\{ c_{nm} \right\}$  reduces to an $N \times N$ submatrix, where the $N$ rows are associated with the positions of the $N$ sources and the $N$ columns with the directions pointing towards the $N$ detectors.

\subsection{$N$-photon Hong-Ou-Mandel interference}

We are interested in the overall probability to measure an $N$-photon coincidence event at the $N$ detectors, i.e., the $N$-th order spatial intensity correlation function 
\begin{equation}
G^{(N)}(\vec{r}_1, \ldots, \vec{r}_N) \sim  \left\langle \opd{b}_1 \ldots \opd{b}_N  \op{b}_N \ldots \op{b}_1 \right\rangle_{\ket{\psi_\text{out}}} \, .
\label{eq:GN}
\end{equation}
Taking into account the transformation coefficients $c_{nm}$ (Eq.~(\ref{eq:cnm2})) one can see that at certain detector positions a completely destructive interference effect can occur, depending on the phase factors $\Delta \varphi_m$. This corresponds to the $N$-photon Hong-Ou-Mandel interference effect, i.e., the completely destructive interference of $N$-photon quantum paths, in free space. 

Indeed, if we have a closer look at the final state of Eq.~(\ref{eq:2output}) we can rewrite it in the following way
\begin{equation}
	\ket{\psi_\text{out}} = \ket{\psi'_\text{out}} + \underbrace{\sum_{\sigma} \prod_{m=1}^{N} c_{\sigma(m)m} \, \opd{b}_m \ket{0}}_{\ket{\psi_\text{coinc.}}} \, ,
	\label{eq:2output2}
\end{equation}
where $\ket{\psi'_\text{out}}$ contains all parts of the state which do not contribute to the $N$-photon coincidence measurement at the $N$ detectors. Only states of the form $\ket{1,...\, ,1}$, where $N$ different modes pointing towards the $N$ different detectors are occupied by a single photon, will lead to a signal in the coincidence measurement and are grouped in the second term $\ket{\psi_\text{coinc.}}$. Here $\sigma$ stands for the permutation applied to the set of elements $\{1,...\, ,N\}$  and the sum runs over all possible permutations, where $\sigma(m)$ denotes the $m$-th element of a certain permutation $\sigma$. From Eqs.~(\ref{eq:GN}) and (\ref{eq:2output2}) we obtain for the coincident $N$-th order spatial intensity correlation function
\begin{equation}
	G^{(N)}(\vec{r}_1, \ldots, \vec{r}_N) \sim  \abs{\sum_{\sigma} \prod_{m=1}^{N} c_{\sigma(m)m}}^2 \, .
	\label{eq:2pcoinc}	
\end{equation}

\begin{figure}
	\centering
	\includegraphics{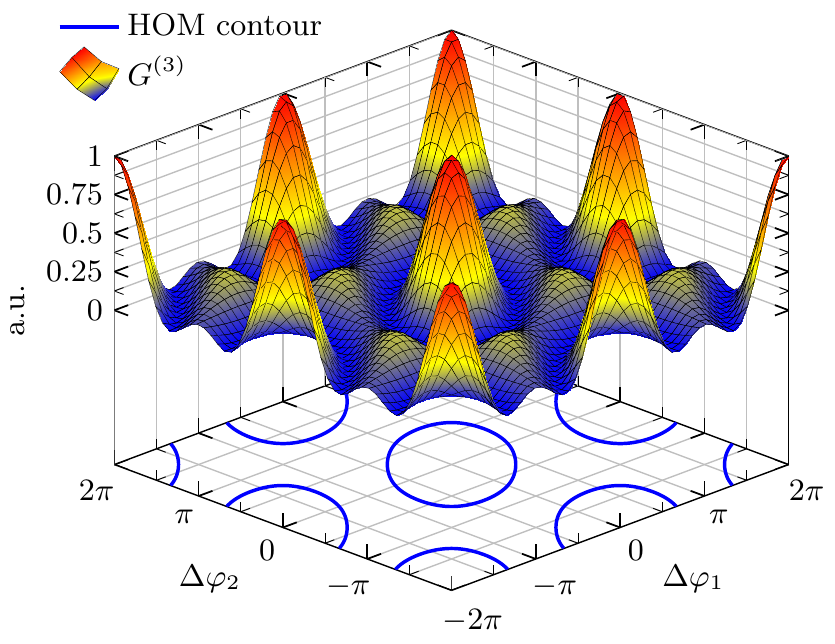}
	\caption{Plot of the third order spatial intensity correlation function $G^{(3)}$ for $N=3$ atoms in free space as a function of $\Delta \varphi_1$ and $\Delta \varphi_2$ for $\Delta \varphi_3=0$. Completely destructive Hong-Ou-Mandel interference is found, if $G^{(3)}=0$. The values for $\Delta \varphi_1$ and $\Delta \varphi_1$ fulfilling this condition are indicated by the contour plot at the bottom.}
	\label{fig:3HOM}
\end{figure}

In order to find the positions, where a completely destructive interference of the $N$-photon quantum paths, i.e., the $N$-photon Hong-Ou-Mandel dip, is observed, we must search for $G^{(N)}(\vec{r}_1, \ldots, \vec{r}_N) \mbeq 0$. Inserting Eq.~(\ref{eq:cnm2}) into Eq.~(\ref{eq:2pcoinc}), we find the following condition

\begin{equation}
	\sum_{\sigma} \prod_{m=1}^{N} \, e^{-\text{i} \, \sigma(m) \cdot \Delta \varphi_m} \mbeq 0 \, ,
	\label{eq:cond}
\end{equation}
where we dropped the normalization factor $c_\text{norm}$. The task is thus to find the right phases $\Delta \varphi_m$, i.e., the right detector positions, to fulfill Eq.~(\ref{eq:cond}). One of the solutions is given by
\begin{equation}
	\begin{split}
		\Delta \varphi_1 &= \frac{2 \pi}{N} \\
		\Delta \varphi_m &= 2 \pi \cdot m \quad \text{for $m\neq1$}.
	\end{split}
	\label{eq:2positions}
\end{equation}
To prove that these positions actually lead to a $N$-photon Hong-Ou-Mandel dip we calculate the expression explicitly and find
\begin{equation}
	\begin{split}
		\sum_{\sigma} \prod_{m=1}^{N} \, e^{-\text{i} \, \sigma(m) \cdot \Delta \varphi_m} &= \sum_{\sigma} e^{-\text{i} \sum_{m=1}^{N} \, \sigma(m) \cdot \Delta \varphi_m} \\
		&= \sum_{\sigma} e^{-\text{i} \cdot \sigma(1) \cdot \Delta \varphi_1} \\
		&= (N-1)! \sum_{m=1}^N e^{-\text{i} \cdot m \cdot \Delta \varphi_1} \\
		&= (N-1)! \sum_{m=1}^N {\left( e^{-\text{i} \cdot \frac{2 \pi}{N}}\right)}^m \\
		&= 0 \, ,
	\end{split}
	\label{eq:2prove}
\end{equation}
where we used the following properties
\begin{itemize}
	\item all phases $\Delta \varphi_m = 2 \pi \cdot m$, for $m\neq1$, do not contribute to the sum in the exponent, as $e^{\text{i} \phi + 2 \pi \cdot j} = e^{\text{i} \phi}$ with $j \in \mathbb{Z}$
	\item for a specific $m$, there are $(N-1)!$ different permutations $\sigma$ for which $\sigma(1)=m$
	\item according to Eq.~(\ref{eq:2positions}), the phase $\Delta \varphi_1$ is defined as the $N$-th root of unity
	\item the sum over all the $N$-th roots of unity is equal to zero (proof via geometric series).
\end{itemize}
Note that in Eq.~(\ref{eq:2positions}) we can not drop the factor $m$ in $\Delta \varphi_m$, although it would be possible from a mathematical point of view, since the problem is $2\pi$-periodic. But in order to perform the experiment, we need $N$ \textit{different}  ``plane wave modes'' as output modes. These modes are determined by the detector positions which in turn determine the phases $\Delta \varphi_m$. Since the detector positions must be physically different from each other, the phases $\Delta \varphi_m$ must be different as well. A look at Eq.~(\ref{phasetodetector}) reveals that this is not a problem, as we can choose for a given $k$ the distance $d$ between the sources appropriately so that $k \cdot d > 2 \pi \cdot N$. Note that the detector configuration for the observation of the $N$-photon Hong-Ou-Mandel effect given in Eq.~(\ref{eq:2positions}) is not the only possible one. There are infinitely more sets of phases, which lead to a completely destructive interference. As an example the third order spatial intensity correlation function $G^{(3)}(\vec{r}_1, \vec{r}_2, \vec{r}_3)$ is shown in Fig.~\ref{fig:3HOM}, where the detector positions $\vec{r}_m$ are expressed as phases $\Delta \varphi_m$ as discussed in Sec. \ref{subsec:Ngeo}. The infinite range of values for $N = 3$ atoms where $G^{(3)}(\Delta \varphi_1, \Delta \varphi_2, \Delta \varphi_3)=0$ is shown as a Hong-Ou-Mandel contour in the bottom of the plot.

\section{Conclusion}

In this paper we proposed a new kind of multi-photon interferometer mixing the different modes not via a set of beam splitters or elaborate photonic circuits but merely by detecting the photons in the far field in free space. As no optical elements are employed the setup represents a most simple and natural realization of a multi-photon interferometer. We first investigated the case of two independent incoherent photons as an input state and demonstrated that the setup allows at certain detector positions to obtain a completely destructive interference in the corresponding two-photon coincidence. This is the analogue to the original Hong-Ou-Mandel experiment, however without the use of a beam splitter. Additionally, we showed that the scheme can be generalized to obtain also for an arbitrary number $N$ of independent incoherent photons as an input state a vanishing $N$-photon coincidence $G^{(N)}(\vec{r}_1, \ldots, \vec{r}_N)$ in case of appropriate detector positions $\vec{r}_1, \ldots, \vec{r}_N$, corresponding to a generalized $N$-photon Hong-Ou-Mandel interference effect in free space.

Note that the measurement of the generalized $N$-photon Hong-Ou-Mandel interference effect in free space is implemented by use of a Hanbury Brown and Twiss type interferometer, where the intensities are coincidentally recorded by detectors at different positions in the far field of the sources and correlated. We thus have established in this paper a close connection between the Hanbury Brown and Twiss interference effect and the Hong-Ou-Mandel effect for $N$ identical statistically independent incoherent single photons.

\begin{acknowledgments}
The authors gratefully acknowledge funding by the chair of the Institute for Theoretical Physics II, Department Physik, Universit\"at Erlangen-N\"urnberg.
\end{acknowledgments}

\bibliography{library}

\end{document}